\documentclass[pdflatex,sn-mathphys]{sn-jnl}
\jyear{2021}%

\theoremstyle{thmstyleone}%
\usepackage{amsmath}
\usepackage{graphicx}
\usepackage{hyperref}

\usepackage{xcolor}

\definecolor{myred}{RGB}{200,0,0}
\definecolor{myblue}{RGB}{0,0,180}

\usepackage[normalem]{ulem}  

\hypersetup{
    colorlinks=true,
    linkcolor=blue,
    filecolor=magenta,      
    urlcolor=cyan,
    bookmarksdepth=4,
}

\theoremstyle{thmstyletwo}%

\theoremstyle{thmstylethree}%

\begin{document}


\title[Article Title]{Self-consistent gravity model for inferring node mass in flow networks}

\author[1,2]{\fnm{Daekyung} \sur{Lee}}

\author[3]{\fnm{Wonguk} \sur{Cho}}

\author[1]{\fnm{Heetae} \sur{Kim}}

\author*[4]{\fnm{Gunn} \sur{Kim}}\email{gunnkim@sejong.ac.kr}

\author[4,5]{\fnm{Hyeong-Chai} \sur{Jeong}}

\author*[6]{\fnm{Beom Jun} \sur{Kim}}\email{beomjun@skku.edu}

\affil[1]{\orgdiv{Department of Energy Engineering}, \orgname{Korea Institute of Energy Technology}, \orgaddress{\city{Naju}, \postcode{58322}, \country{Republic of Korea}}}

\affil[2]{\orgname{Supply Chain Intelligence Institute Austria}, \orgaddress{\city{Vienna}, \postcode{1030}, \country{Austria}}}

\affil[3]{\orgdiv{Graduate School of Data Science}, \orgname{Seoul National University}, \orgaddress{ \city{Seoul}, \postcode{08826}, \country{Republic of Korea}}}

\affil[4]{\orgdiv{Department of Physics and Astronomy \& Institute for Fundamental Physics}, \orgname{Sejong University}, \orgaddress{\city{Seoul}, \postcode{05006}, \country{Republic of Korea}}}

\affil[5]{\orgdiv{School of Computational Sciences}, \orgname{Korea Institute for Advanced Study}, 
\orgaddress{\city{Seoul}, \postcode{02455}, \country{Republic of Korea}}}

\affil[6]{\orgdiv{Department of Physics}, \orgname{Sungkyunkwan University}, \orgaddress{ \city{Suwon}, \postcode{16419}, \country{Republic of Korea}}}

\newcommand{\noteDK}[1]{\textbf{\textcolor{red}{#1}}}


\abstract{The gravity model, inspired by Newton's law of universal gravitation, has been a cornerstone in the analysis of trade flows between countries. In this model, each country is assigned an economic mass, where greater economic masses lead to stronger trade interactions. Traditionally, proxy variables like gross domestic product (GDP) or other economic indicators have been used to approximate this economic mass. While these proxies offer convenient estimates of a country's economic size, they lack a direct theoretical connection to the actual drivers of trade flows, potentially leading to inconsistencies and misinterpretations. To address these limitations, we present a data-driven, self-consistent numerical approach that infers economic mass directly from trade flow data, eliminating the need for arbitrary proxies. Our approach, tested on synthetic data, accurately reconstructs predefined embeddings and system attributes, demonstrating robust predictive accuracy and flexibility. When applied to real-world trade networks, our method not only captures trade flows with precision but also distinguishes a country’s intrinsic trade capacity from external factors, providing clearer insights into the key elements shaping the global trade landscape. This study marks a significant shift in the application of the gravity model, offering a more comprehensive tool for analyzing complex systems and revealing new insights across various fields, including global trade, traffic engineering, epidemic prevention, and infrastructure design.}




\maketitle

\section{Introduction}\label{sec1}

Natural phenomena have long inspired theories aimed at understanding human behavior and social dynamics. Among these, the gravity model stands out for its analogy between Newton's law of universal gravitation and trade flows between countries\cite{gravity_start, thegravity}. This model posits that the volume of trade between two countries is proportional to their economic 'mass' and inversely related to the distance between them. Over the past seventy years, this intuitive model has become a cornerstone in the study of international trade\cite{trade1, trade2, trade3, trade4, trade5, trade6, trade7, trade8, trade9, trade10, trade11}. While early empirical applications were criticized for lacking a solid theoretical foundation, subsequent theoretical advances have not only strengthened the model's basis\cite{strength, uniqueness, border, error, gravity_base} but also extended its applicability to a range of phenomena\cite{demographic, citation, epidemic, epidemic2, mobility}.

Despite its widespread use and theoretical refinement, significant ambiguity remains in the gravity model: the definition and measurement of 'mass'. In the absence of a method to quantify mass directly, most studies rely on proxies form external datasets. However, this reliance on approximate proxies undermines the fundamental consistency of the gravity model, limiting its ability to accurately describe and predict complex human dynamics. For instance, GDP, which is commonly used as a proxy for economic mass in international trade\cite{trade1,trade2,trade3, trade4,trade5}, has been widely criticized for representing only the aggregate market activity of a country while failing to distinguish between costs and benefits, account for structural trade potential, or reflect institutional and compositional factors\cite{GDP_limit}. These limitations cast doubt on its validity as a measure of a country's capacity to export or import within the global network. Similar limitations arise in other domains as well. External indicators are commonly used to represent node attributes in transportation\cite{highway, bus, urban1, urban2, urban3, urban4, urban5}, finance\cite{finance1, finance2, finance3, finance4, finance5}, and other social networks\cite{social1, social2, social3, social4, social5}. While such proxies may yield some descriptive or empirical insight, rigorously validating whether they accurately represent a node’s capacity to generate and attract flow remains a highly challenging task.

In this paper, we present a novel {\em self-consistent} numerical approach to the gravity model that differs fundamentally from traditional methods. While previous studies treat mass as an independent variable introduced to model observed flows, we instead formulate mass as an internal parameter that is inferred together with the deterrence function in order to best reconstruct the observed flow network. This innovative methodology not only improves the consistency and accuracy of the gravity model but also redefines mass as an intrinsic metric reflecting each entity's inherent ability to generate or attract flow. Our validation on synthetic networks demonstrates exceptional accuracy in inferring predefined mass distributions and spatial dependencies, achieving near-perfect precision.

Furthermore, we apply this methodology to the international trade network, yielding a more accurate inference of the system's spatial dependency and outperforming previous methods in reconstructing original trade flows. Our newly defined economic mass metric allows for a nuanced decoupling of each country's intrinsic trade capacity from the market effects of surrounding countries, based on its total export/import data. This refined analysis provides a detailed view of the global trade landscape, particularly emphasizing the significant influence of economic superpowers: the US, China, and Germany.

By advancing our understanding of the gravity model and its empirical application, our approach not only contributes to developing a perspective on the profound structure of global trade but also holds significant potential to refine practices in related fields. While our study focuses on international trade, the implications of this self-consistent numerical approach extend far beyond. The methodology we propose is of considerable value in any field where gravity theory has been applied, including transportation, finance, epidemiology, and social network analysis. By providing a more accurate and consistent framework for understanding and predicting complex systems, our approach promises to substantially advance research and practical applications across a wide range of disciplines.

\section{Result}
\subsection{Limitations of Previous Approaches}\label{sec2}
In the conventional framework of the gravity model, the directed flow $f_{ij}$ from region $i$ to region $j$ is formulated as 
\begin{equation}\label{eq:fij}
f_{ij} = m^{\rm out}_{i}m^{\rm in}_{j}Q(d_{ij}), 
\end{equation}
where $m^{\rm out}_{i}$ and $m^{\rm in}_{j}$ denote the outward and inward {\em mass} of regions $i$, respectively, representing their abilities to generate or attract flows. The function $Q(d_{ij})$, known as the deterrence function, captures the effect of the distance $d_{ij}$ between regions $i$ and $j$ on their interaction strength. It is important to note that the variables in this formulation fall into two categories: $f_{ij}$ and $d_{ij}$ are measurable variables obtained from data, while $m^{\rm out}_{i}$, $m^{\rm in}_{i}$ and $Q(d_{ij})$ are latent variables that describe the underlying structure of the system. Consequently, the primary objective of the gravity model is to infer the values of $m^{\rm out}_{i}, m^{\rm in}_{i}$ and $Q(d_{ij})$ based on the available flow data $ \{ f_{ij} \}$ and distances $\{ d_{ij} \}$.

However, inferring the gravity model's latent variables poses a significant challenge. Since the mass distribution and the deterrence function reflect distinct aspects of the flow data, simultaneously inferring both variables has been considered infeasible without additional information. As an alternative, existing approaches often approximate each node's mass using external indicators, such as GDP for countries in trade networks\cite{trade1,trade2}, or population for locations in transportation networks \cite{bus, china, highway}. While these methods offer useful approximations, they depend on external data sources that cannot guarantee an accurate representation of each node’s capacity to generate and attract flow.

To overcome this limitation, another widely used approach is the strength approximation proposed by L. S. Martyn\cite{strength}. In this model, the inward flow $S_{i}^{\rm in}=\sum_j f_{ji}$ and outward flow $S_i^{\rm out} = \sum_j f_{ij}$ of each node are taken as proxies for $m_i^{in}$ and $m_i^{\rm out}$, respectively. This approach has the merit of utilizing the flow data itself to estimate the masses, without relying on external datasets.

However, the strength approximation still cannot properly represent the mass in the gravity model. The core issue lies in the fact that strength is not an intrinsic property of a node but depends on the entire network. In the gravity model, the mass of node $i$ is an inherent characteristic, determined solely by $i$ itself. In contrast, the strength, by definition, depends on flows involving other nodes. If the masses of another node $j$ changes, the flow $f_{ij}$ and $f_{ji}$, and consequently the strength of node $i$, may change, even if $i$'s own properties remain constant. This dependency implies that strength cannot be considered an intrinsic quantity of the node, thereby limiting its effectiveness in accurately representing the masses in the gravity model. Despite the progress and widespread applications of the strength approximation and other methods, a conclusive resolution has yet to be achieved, leaving the accurate depiction of mass an unresolved challenge. (Further details on the previous gravity model are provided in Supplementary Section 1.)

\begin{figure}[ht]
\centering
\includegraphics[width = \textwidth]{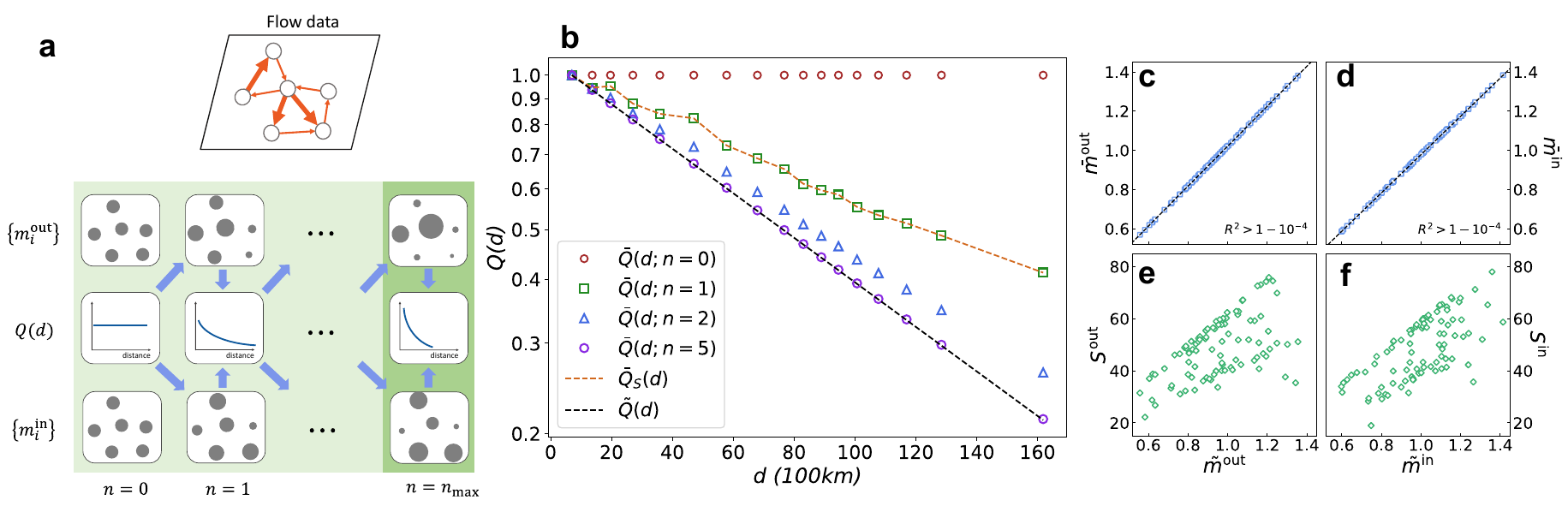}
\caption{The conceptual description of the self-consistent inference algorithm and its verification in model network. (a) The mass distribution $\{ m_{i}^{\rm in, out}\}$ and deterrence function $Q(d)$ are properly initialized in $n=0$ and iteratively update each other to explain the flow dynamics of the system. (b) In the synthetic model network, the inference of deterrence function $\bar{Q}(d,n)$ quickly converges to the ground truth $\tilde{Q}(d)$, while that of strength approximation $\bar{Q}_{S}(d)$ give a rather inaccurate result. [c - d] Comparison of synthetic input mass information $\{ {\tilde m}_i^{\rm out}, {\tilde m}_i^{\rm in} \}$ and the inferred information $\{ {\bar m}_i^{\rm out}, {\bar m}_i^{\rm in} \}$, respectively, from 
our self-consistent formulation of the gravity model. [e - f] Comparison of $\{ {\tilde m}_i^{\rm out}, {\tilde m}_i^{\rm in} \}$ with the estimations from the simple strength approximation in which inward and outward mass distribution are approximated as $\{ S_i^{\rm out}, S_i^{\rm in} \}$.
Our self-consistent inference formulation can clearly reconstruct the information underlying synthetic data of $f_{ij} 
 $ while the strength-based simple gravity model fails.}
\label{fig:concept}
\end{figure}

\subsection{Self-consistent inference formulation}

In response to these challenges, we introduce a novel algorithm specifically designed for gravity model inference. Our method accurately identifies the latent variable $\{m^{\rm out}_{i}, m^{\rm in}_{i}\}$, and $Q(d)$ through a fully self-consistent approach without the need for external data. The concept of our methodology is inspired by certain established techniques\cite{strength}, previously developed for inferring one of the two key variables when the other is available. Specifically, if the system's mass distribution is obtained through an external proxy, these techniques can determine the deterrence function $Q(d)$, and \textit{vice versa}. Although those techniques alone still require external information, we have discovered that integrating them creates a self-consistent loop where the mass distribution and deterrence function alternately refine each other. Starting with simple initial conditions for the latent variables, our approach implements an iterative scheme that sequentially updates each component in alignment with the observed flow data $\{ f_{ij}\} $. We remark that our model reinterprets the gravity model not as a regression-based explanatory framework, but as an inference-based approach that identifies internal embedded variables which best reproduce the observed flow network. This perspective reinterprets mass not as an independent explanatory factor, but as a structurally embedded variable recovered through system-level consistency.

The proposed formulation is graphically depicted in Fig.~\ref{fig:concept} (a). At the initial stage $n=0$, the $\{ m^{\rm out}_{i}, m^{\rm in}_{i} \}$ and $Q(d)$ are initialized as uniform distribution and constant function, which do not yet reflect the $\{ f_{ij} \}$. Moving to stage $n=1$, each node's mass is recalibrated based on the $Q(d)$ from the prior stage to better represent its connected flows; nodes associated with larger outward flows in the actual data are assigned greater outward mass, while those with significant inward flows are attributed a greater inward mass. Following this, the spatial dependency $Q(d)$ is updated based on the revised node mass distribution. Those iterative refinements enhance the model's alignment with the actual flow dynamics $\{ f_{ij} \}$ and continue until all inferred information converges to a stationary state. We explain the detailed procedure of our methodology in the Methods section.

\subsection{Verification with synthetic data}
In this section, we apply our inference formulation to flow data to evaluate its performance and conduct a comparative analysis with existing gravity model methodologies. Our objective is to assess whether this formulation can accurately identify the mass distribution and the deterrence function. Given the absence of known mass distribution and deterrence function in real-world data, we turn to synthetic flow data generated based on arbitrarily assigned mass and deterrence functions. This synthetic data is not intended to replicate empirical trade patterns, but rather to test whether the algorithm can successfully recover the underlying parameters solely from flow information. If the outcomes from our algorithm align with these initial assumptions, it would indicate our method's effectiveness.

To begin, we specify arbitrary values for the mass distribution and deterrence function, denoted as $ \{ {\tilde m}_i^{\rm out}$, ${\tilde m}_i^{\rm in}$ \} and ${\tilde Q}(d)$. These variables enable us to compute the flow data $\{ f_{ij} \}$, following Eq.(\ref{eq:fij}). This generated data then acts as the input for our self-consistent formulation. We apply our inference technique to extract the inferred mass distribution ${\bar m}_i^{\rm out}$, ${\bar m}_i^{\rm in}$, and the inferred deterrence function ${\bar Q}(d)$. The accuracy of our methodology is determined by comparing these inferred quantities with the original variables, serving as our benchmark for success.

We use synthetic data based on an international trade network spanning 94 countries as a backbone structure. In this simulation, mass variable $\tilde{m}_{i}^{\rm out}$ and $\tilde{m}_{i}^{\rm in}$ are allocated randomly following a normal distribution with a mean of 1 and a standard deviation of 0.2. The deterrence function is assigned as $\tilde{Q}(d) = e^{-d/10000}$, representing one of its typical forms from previous research. The distance $d_{ij}$ between a node pair is defined as the geodesic distance, in kilometers, between the reference locations of the two countries measured along a great circle on the earth. We assign each country's reference location from the Google Dataset Publishing Language\cite{location}. With these latent and real variables, we calculate the flows $f_{ij}$ according to Eq.~(\ref{eq:fij}), providing a benchmark for our model's performance.

Our gravity model initiates the inference process with a set of arbitrary initial values, iteratively updating the mass distribution and deterrence function. This iterative cycle continues until the inferred values—${\bar m}_i^{\rm out}$, ${\bar m}_i^{\rm in}$, and ${\bar Q}(d)$—no longer exhibit significant changes, indicating that a steady state has been reached. It is important to note that these values include various scaling factors that require careful adjustment for accuracy. To ensure a fair comparison, we normalize the maximum value of the deterrence function to unity and adjust the averages of inward mass and outward mass distributions to be equal.

In Fig.~\ref{fig:concept} (b), we demonstrate the iterative refinement of the inferred deterrence function ${\bar Q}(d; n)$, highlighting how it gradually aligns with the synthetic baseline ${\tilde Q}(d)$ over iterations $n=1, 2$, and 5. Starting as a constant at $n=0$, ${\bar Q}(d; n)$ undergoes adjustments at each step, marked distinctly for $n=0, 1, 2, 5$. Despite limitations from piece-wise linear approximation and the binning process, by iteration $n=5$, ${\bar Q}(d; n)$ aligns closely with ${\tilde Q}(d)$, depicted by the black dashed line representing our model's ground truth.

Furthermore, we compare this progression with the deterrence function ${\bar
Q}^S(d)$ derived from the strength approximation, calculated as ${\bar
Q}^{S}(d) = \langle f_{ij} / S_{i}^{\rm out}S_{j}^{\rm in} \rangle_{d \approx
d_{ij}}$, applying the same binning strategy. Interestingly, after the first
iteration, ${\bar Q}(d; n=1)$ closely mirrors ${\bar Q}^S(d)$. We note that
this resemblance is attributed to the gravity model's intrinsic property.
Specifically, under a constant $Q(d)$, the outward strength $S_i^{\rm
out}$—expressed as $\sum_j f_{ij} = m_i^{\rm out} \sum_{j} A_{ij} m_j^{\rm in}
Q(d_{ij})$ with $A_{ij}$ being the element of the directed adjacency matrix
—becomes proportional to $m_i^{\rm out}$, since the summation across
$j$ is approximately equivalent for all nodes $i$. This rationale similarly
applies to the inward strength, making the deterrence function of the first
iteration resemble ${\bar Q}^{S}(d)$. As we advance beyond the initial
iteration, our model quickly diverges from these early parallels and converges
towards ${\tilde Q}(d)$, the synthetic benchmark. This progression underscores
our method's capability to dynamically refine and accurately model the
deterrence function within a synthetic network framework.

In Figs.~\ref{fig:concept} (c)-(f), we compare the synthetic input mass distribution ${{\tilde m}_i^{\rm out}, {\tilde m}_i^{\rm in} }$ with the inferred values from our model and the strength approximation. Data points are plotted with the ground truth on the horizontal axis and the inferred masses and strengths on the vertical. Notably, in panels (c) and (d), the masses inferred by our model exhibit a tight correlation with the synthetic values, as demonstrated by a linear fit characterized by a high coefficient of determination. Conversely, the results from the strength approximation, depicted in panels (e) and (f), reveal a marked divergence from the synthetic benchmarks. This underscores the strength approximation's limitations in accurately capturing the network's intrinsic mass distribution.

\begin{figure}[ht]
\centering
\includegraphics[width = 0.9\textwidth]{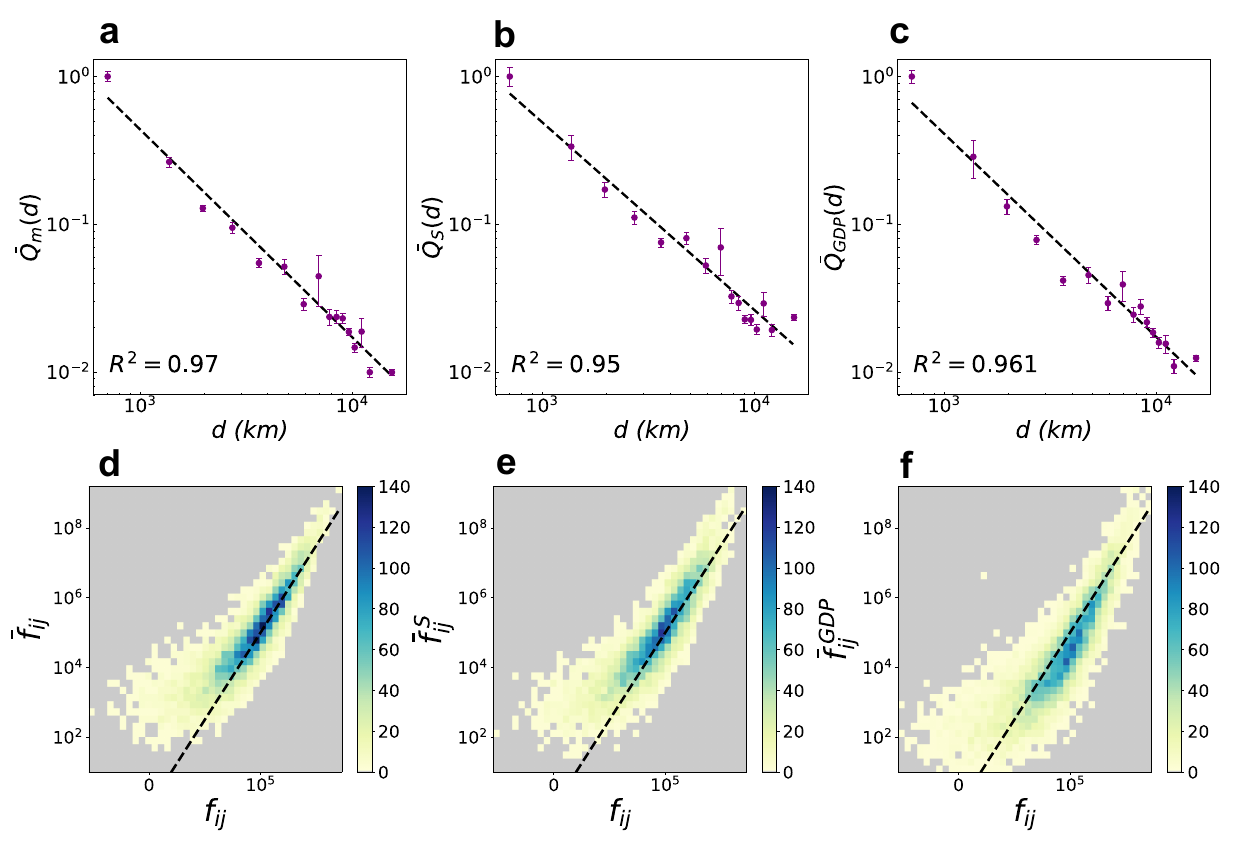}
\caption{[(a) - (c)] The deterrence functions ${\bar Q}m(d)$, ${\bar Q}^{S}(d)$, and ${\bar Q}^{\rm GDP}(d)$ for the 2019 international trade network, estimated using (a) our method, (b) the strength approximation, and (c) the GDP approximation, respectively. Although all functions follow a power-law decay pattern, the results from our method (a) feature narrower error bars and a more uniform decay pattern. [(d) - (f)] Density plots compare the actual trade flows, denoted by $f_{ij}$, with those predicted by the three methodologies. These include (d) our method with ${\bar f}_{ij}$, (e) the strength approximation with ${\bar f}^{S}_{ij}$, and (f) the GDP approximation with ${\bar f}_{ij}^{\rm GDP}$. The plots visually underscore our method's better ability to reconstruct the real flow data, particularly in regions of high trade flow.}
\label{fig:real_result}
\end{figure}

\subsection{International trade network analysis}

In this section, we explore the practical applications of our gravity
formulation by analyzing real-world data. Specifically, we use the
international trade network as our testbed, which closely aligns with the
gravity model's primary focus. The 2019 international trade network data, sourced from the BACI dataset provided by the Centre d'Études Prospectives et d'Informations Internationales (CEPII)~\cite{BACI, cho2023multiresolution}. In this
network, each node $i$ represents a country and a weighted directed edge
$f_{ij}$ denotes the total annual trade flow from country $i$ to country $j$.  The distance $d_{ij}$ is again defined as the geodesic distance.
Note that the BACI dataset includes over 200 countries; however, we use only the 94 countries included in all datasets for consistency with other datasets introduced later. The list of these countries can be found in the Supplementary Information.

Employing our method, we estimate the mass distribution ${ {\bar m}_i^{\rm out}, {\bar m}_i^{\rm in} }$ and the deterrence function ${\bar Q}(d)$, which stabilizes after twenty iterations ($n=20$). We also use conventional strength and GDP-based approximations to estimate a country's mass. The GDP of each country is obtained from The World Bank database\cite{worldbank}, including only countries with a GDP higher than 100 million dollars. We adjust each variable's scale to normalize the deterrence function's maximum to unity and to balance the mean values of the inward and outward mass distributions. In Fig.~\ref{fig:real_result} (a)-(c), the derived deterrence functions are displayed: (a) ${\bar Q}_m(d)$ from our method, (b) ${\bar Q}^S(d)$ from the strength approximation, and (c) ${\bar Q}^{\rm GDP}(d)$ from the GDP approximation. We also plotted the linear regression line in log space for each dataset, represented by the black dashed line. Our methodology results in smaller error bars and a higher coefficient of determination ($R^2$) for the linear regression line in (a), indicating superior accuracy and consistency compared to the alternative approximations.

As a further validation, we reconstruct the flow distribution $\{ {\bar f}_{ij} \}$ by utilizing the inferred masses $\{ {\bar m}_i^{\rm out}, {\bar m}_i^{\rm in} \}$ and the deterrence function ${\bar Q}(d)$ from various methods. To ensure consistency, we adjust the scale of the reconstructed flow distribution to match the total flow sum of the actual data. The reconstruction is compared to the actual flow distribution $\{ f_{ij} \}$, as illustrated in the density plots shown in  Fig.~\ref{fig:real_result} (d)-(f), where the horizontal and vertical axes denote the actual and reconstructed flows, respectively. We draw a black dashed diagonal line in each plot, indicating perfect agreement. Our method's superiority is evident as it shows the highest density of points near the black dashed diagonal line. We note that the goodness-of-fit measured by the Sørensen–Dice similarity index (SSI) is significantly higher for our method, as shown in Supplementary Fig.~S3. Although our method slightly overestimates the flow values for smaller $f_{ij}$, it achieves significantly higher accuracy across most data points than the strength and GDP-based approximations. This highlights the effectiveness of our method in capturing the dynamics of international trade flow. Given that real-world data are often incomplete, we tested our method's ability to generalize by predicting missing flows from a partial dataset. This out-of-sample validation helps address concerns about overfitting and demonstrates the robustness of our approach. The results are presented in Supplementary Fig.~S3.
\begin{figure}[ht]
    \centering
    \includegraphics[width=0.9\textwidth]{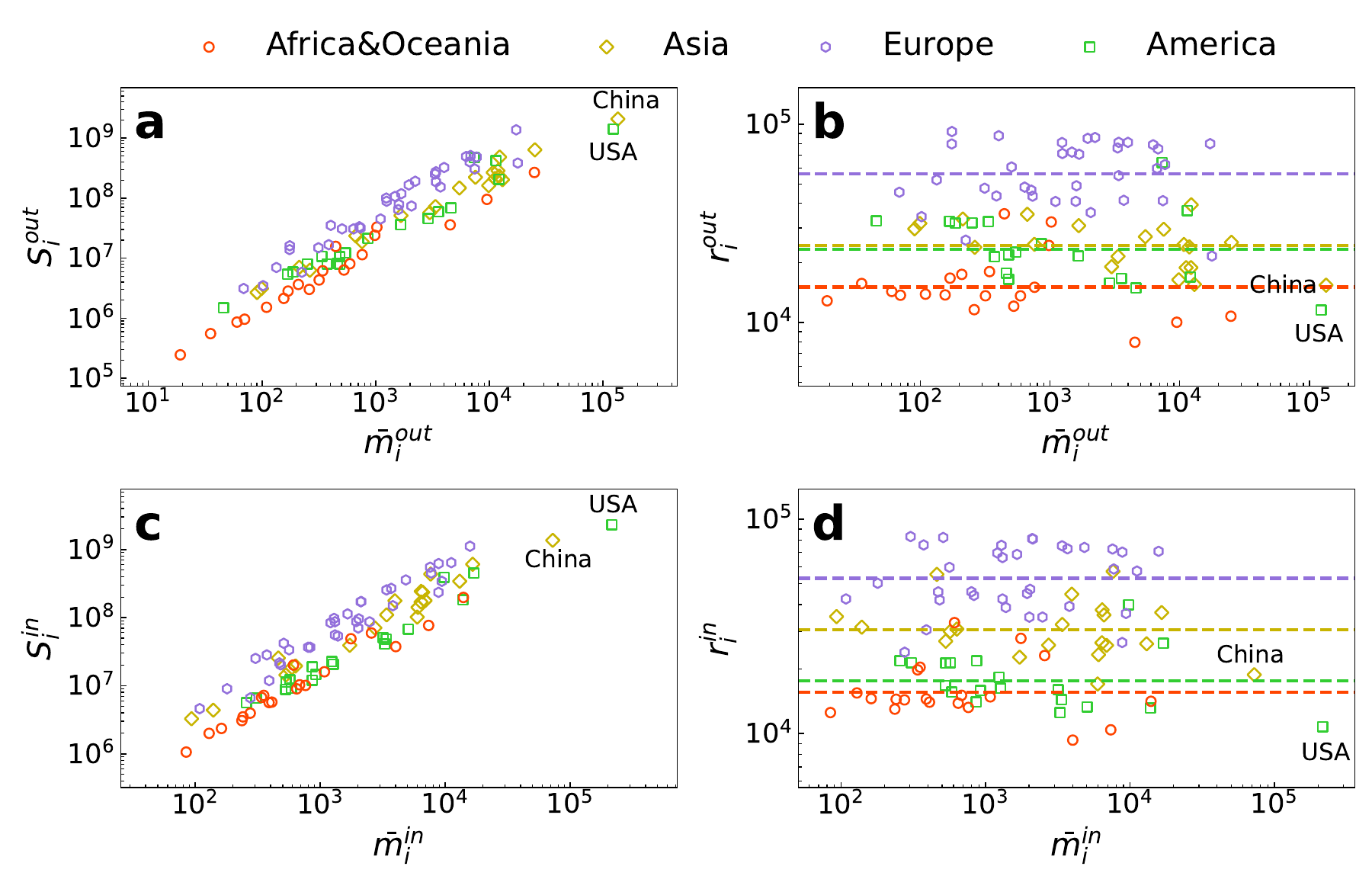}
    \caption{Comparison of inferred mass and strength distributions for the international trade network. Symbols represent each country's mass and strength in each panel, classified by four continent labels. The left panels (a) and (c) illustrate the relationship between strength and mass, whereas the right panels (b) and (d) display the External Advantage Index (EAI), which represents the ratio of strength to mass for each node. The positioning of data points in the left panels is noticeably grouped according to their continent label, suggesting the geographical implications of the strength-mass ratio. The results in the right panels further emphasize these patterns in the EAI, with colored dashed lines indicating the geometric average of data points for each continent.}
    \label{fig:ITN_MS}
\end{figure}

The inferred mass distributions at iteration $n=20$ are illustrated in Fig.~\ref{fig:ITN_MS}, and compared with the strength values using log-log scale scatter plots due to the data distribution (see Supplementary Fig.~S1). This comparison is conducted separately for outward and inward quantities, presented in the top and bottom panels. Upon inspection, Fig.~\ref{fig:ITN_MS}(a) exposes a prominent pattern: the majority of European counties (violet data points) exhibit a tendency toward relatively high bidirectional strength values, $S_{i}^{\rm in, out}$, in relation to the corresponding outward mass inference values, ${\bar m}_{i}^{\rm in, out}$. Conversely, most African and Oceanian countries (red data points) show the opposite trend. This pattern is consistent for both outward and inward quantities, as shown in Fig.~\ref{fig:ITN_MS}(c).

We propose that the behaviors observed in Fig.~\ref{fig:ITN_MS} can be attributed to the inherent characteristics of the gravity model. Specifically, according to Eq.(\ref{eq:fij}), the outward strength of node $i$ is described as $S_i^{\rm out} = \sum_{j} f_{ij} = {\bar m}_i^{\rm out} \sum_{j} A_{ij}{\bar m}_{j}^{\rm in} {\bar Q}(d_{ij})$. This formulation reveals that a node's outward strength is determined not solely by its own outward mass but also by the weighted sum of the inward masses of surrounding nodes. Consequently, the ratio 
\begin{equation}\label{eq:EAI}
r_{i}^{\rm out} = \frac{S_i^{\rm out}}{{\bar m}_i^{\rm out}} = \sum_{j} A_{ij}{\bar m}_{j}^{\rm in} {\bar Q}(d_{ij})
\end{equation}
serves as an indicator of external contributions to node $i$'s total exports, which we term the External Advantage Index (EAI). Nodes surrounded by others with substantial inward masses are likely to exhibit a high $r_i^{\rm out}$, signifying a considerable export advantage due to their neighbors. Similarly, this logic applies to inward strength, $S_i^{\rm in}$, which is expressed as the product of its inward mass and the weighted sum of surrounding nodes' outward masses, further illustrating the reciprocal nature of trade relationships.

This concept aligns with the observations from Fig.~\ref{fig:ITN_MS}(a) and (c). European countries, characterized by their dense geographical distribution and substantial masses, tend to exhibit higher EAI values compared to the countries in Africa and Oceania. To further highlight the continent-specific characteristics of global trade, panels (b) and (d) plot the bidirectional EAI against the mass of each node, with colored lines representing the geometric average EAI for each continent. The geometric average is employed to accurately represent the tendency of data points distributed along log scale axes. The average EAI for European countries (violet) is notably highest, followed by those in Asia, the Americas, Africa $\&$ Oceania, in decreasing order.

\begin{figure}[ht]
    \centering
    \includegraphics[width=1.0\linewidth]{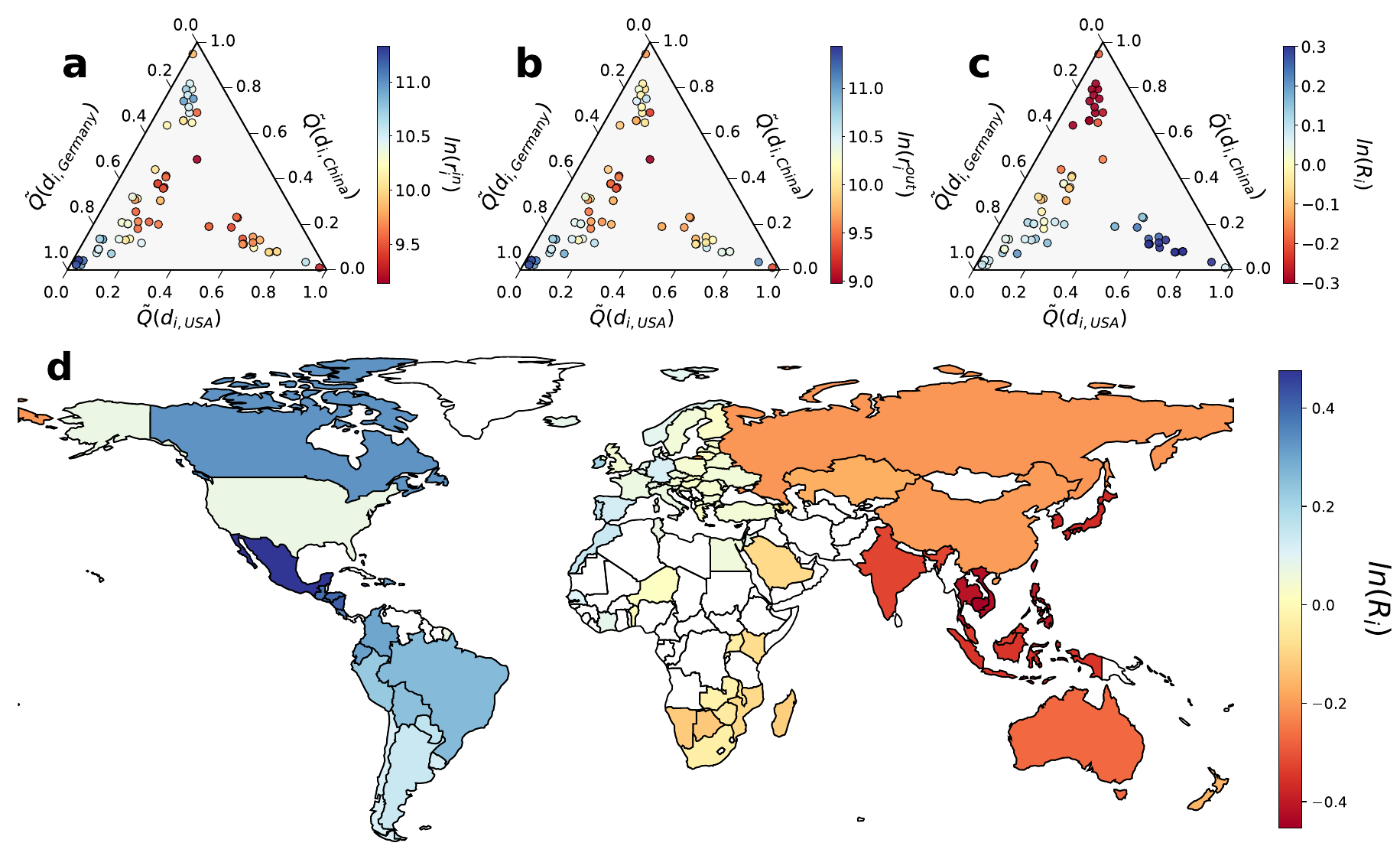}
    \caption{
The ternary plot illustrates the trade patterns of countries, arranged by their relative deterrence functions towards three major countries: Germany, the USA, and China. A logarithmic color scale is utilized to enhance the visualization of specific details. Panels (a) and (b) depict the External Advantage Index (EAI) for imports and exports, respectively. The analysis reveals that proximity to China and Germany correlates with advantages in importation and exportation, while proximity to the USA is mainly associated with export benefits. Panel (c) presents the Relative Advantage Index (RAI), defined as the ratio between outward and inward EAI, reflecting the economic characteristics of countries within the global trade network. This visualization effectively clusters data points according to the proximity of their closest major country. Furthermore, panel (d) illustrates the geographic distribution of RAI, providing insights into the spatial trade advantage landscape.}
    \label{fig:ternary}
\end{figure}

\begin{table}[ht]
\begin{center}
\begin{tabular}{ |c||c|c|c|c|c|c|c|  }
\hline
 Country & $S^{\rm out}$ & $m^{\rm out}$ & $S^{\rm in}$ & $m^{\rm in}$ & $r^{\rm out}$ & $r^{\rm in}$ & $m^{\rm in}/m^{\rm out}$\\
 \hline
 USA & 1420000 & 123000 & \textbf{2317000} & \textbf{215500} & 11.54 & 10.75 & \textbf{1.753} \\
 Germany & 1372000 & 17180 & 1116000 & 15760 & \textbf{79.87} & \textbf{70.85} & 0.9171 \\
 China & \textbf{2080000} & \textbf{134700} & 1362000 & 72320 & 15.45 & 18.83 & 0.5369 \\
 \hline
\end{tabular}
\end{center}
\caption{Comparison of bidirectional mass and strength, and several applied quantities of three major countries: the USA, Germany, and China. We set the unit of strength and EAI as a billion dollars, and other quantities are unitless. The first four columns reveal that China and the USA have their advantage in exportation and importation, while those of Germany are relatively even. The next two columns exhibit the bidirectional EAI, and the last column denotes the ratio between inward and outward masses. All values in the table are presented with a precision of four significant figures.}
\label{tab:major}
\end{table}

Building on our findings about the mass and EAI distributions, we further explore the global trade landscape involving major economic powers. Given that a country's EAI is influenced by the economic mass of its neighbors, we hypothesize that the economic conditions of countries proximate to dominant trade powers are significantly shaped by these giants' economic scales. For instance, countries adjacent to major economic powers likely experience advantages in their import and export activities. 

To analyze how these advantages manifest in detail, we examined the attributes of each country based on their proximity to dominant economic powers. Specifically, we selected the USA, Germany, and China as representatives of three major economic regions: North America, Europe, and East Asia, respectively. Figure~\ref{fig:ternary} displays each country's inward and outward EAI ($r_{i}^{\rm in, out}$) and their ratio ($R_i = r_{i}^{\rm out}/r_{i}^{\rm in}$) using a ternary plot. For geographic representation, each point on the plot is positioned according to deterrence functions with respect to three major economic powers: the USA, Germany, and China. Consequently, each dominant country is denoted by a data point at its corresponding corner, and others are positioned based on proximity to these powers. We opted for logarithmic scale color mapping in all plots to illustrate the variable patterns effectively. The outcomes depicted in Figs.~\ref{fig:ternary} (a) and (b) mostly concur with our hypotheses: the bidirectional EAI of each data point mainly increases as they approach the corners of plots due to each dominant country's substantial contribution to nearby countries' trade environments. (We also include the geographic distribution of $r_i^{\rm in, out}$ in Supplementary Fig.~S2.)

In Fig.~\ref{fig:ternary} (c), the ternary plot illustrates the out/in EAI ratio, which we term the Relative Advantage Index (RAI). It encapsulates the local economic context of country $i$ regarding its trade balance, highlighting the relative advantage of a country's exports over its imports. In this plot, pronounced clustering is evident, with each data point closely aligning with the economic influence of its nearest major country. Specifically, countries close to the USA and China show marked advantages in exports and imports, respectively, while those near Germany display balanced advantages in both sectors. This pattern of relative trade advantages is further emphasized in Fig.~\ref{fig:ternary} (d), which displays the geographic distribution of RAI and reveals distinct segregation among countries in the global trade landscape. 

To elucidate the observed trend, we analytically assess the influence of dominant countries on their neighbors. Specifically, for a given country $i$, when a dominant country $p$ exhibits substantially higher $\bar{m}_{p}^{\rm in, out}$ and $Q(d_{ip})$ values compared to $i$'s other neighbors, we express the RAI relative to the dominant country's metrics as
\begin{equation}
R_i = \frac{r_i^{\rm out}}{r_{i}^{\rm in}} = \frac{\sum_{j} A_{ij} {\bar m}_{j}^{\rm in} {\bar Q}(d_{ij})}{\sum_{j } A_{ji} {\bar m}_{j}^{\rm out} {\bar Q}(d_{ji})} \approx \frac{{\bar m}_{p}^{\rm in}}{{\bar m}_{p}^{\rm out}}.
\end{equation}
This equation shows that a country's RAI ($R_i$) is heavily influenced by the trade characteristics of its dominant neighbor. Specifically, if the neighbor primarily focuses on exports, the surrounding countries are likely to see increased benefits from imports. Conversely, if the neighbor is mainly import-oriented, it may boost export opportunities for these countries. In other words, the dominant countries' intrinsic economic characteristics significantly influence neighboring countries' trade environments, further shaping the landscape of the global trade network.

The pronounced clustering behavior of the $R_i$ distribution in Fig.~\ref{fig:ternary} (c) strongly supports our interpretation, with each data point's value closely aligning with the inward-to-outward mass ratio of its nearest dominant country: $\frac{m_{\rm USA}^{\rm in}}{m_{\rm USA}^{\rm out}} > 1$, $\frac{m_{\rm Germany}^{\rm in}}{m_{\rm Germany}^{\rm out}} \approx 1$, and $\frac{m_{\rm China}^{\rm in}}{m_{\rm China}^{\rm out}} < 1$ as detailed in Table. \ref{tab:major}. The geographic distribution of $R_{i}$, as depicted in Fig.~\ref{fig:ternary} (d), further corroborates this clustering behavior, underscoring the significant influence of these major economies.

Building on these insights, we can infer that the geographical landscape of international trade is profoundly shaped by the economic characteristics of these three major countries. The United States' import-centric economy affects the export conditions of its neighboring countries. In contrast, China's export-oriented economy allows its nearby countries to enhance their total importation. Similarly, Germany's relatively balanced economy enables neighboring countries to optimize their trade capabilities in both sectors. To support a more detailed understanding of these dynamics, Supplementary Table S1 provides the GDP, export, import, and both inward and outward mass values for each country.

\section{Discussion}\label{sec12}

We have developed a self-consistent inference formulation for gravity models that overcomes the fundamental limitations of previous approaches. Unlike traditional methods that rely on external proxies or approximations for mass, our method directly utilizes flow and distance data to simultaneously infer the mass distribution and the deterrence function. Starting with initial assumptions about mass and deterrence, our iterative refinement process relies solely on the inherent data, eliminating the need for supplementary external information.

This novel framework not only enables the extraction of significant model parameters directly from raw data but also significantly improves the model's accuracy and applicability. Through numerical simulations with synthetic datasets, we have demonstrated that our approach outperforms conventional gravity modeling methods, such as the strength approximation, in accurately identifying the ground truth of mass and the deterrence function. Specifically, our method achieved near-perfect precision in inferring predefined mass distributions and spatial dependencies, highlighting its robustness and effectiveness.

Applying our model to real-world international trade networks has provided profound insights into the global trade landscape. By comparing each country's inferred mass with its total exports and imports, our framework successfully decouples total trade volume into intrinsic capability and external influences. Specifically, our findings indicate that the strength-to-mass ratios capture each country's local economic advantage, which is influenced by the mass distributions and proximity of neighboring countries.

Furthermore, our analysis reveals that the economic advantages of individual countries are significantly shaped by the economic characteristics of dominant countries like the United States, China, and Germany. Countries located near these economic superpowers—characterized by distinct export- or import-oriented economic structures—derive notable benefits in terms of their own trade dynamics. This observation allows us to reinterpret the global trade network landscape, highlighting how it is structured around these dominant countries and how their economic characteristics mold the trade patterns and economic conditions of neighboring nations. Furthermore, future work may explore more realistic definitions of distance, incorporating data such as shipping routes or cultural proximity. An alternative direction would be to extend our framework to infer distances directly from flow data by embedding countries into a latent geometric space, thereby treating distance itself as a structural variable within the model.

Our formulation introduces a novel metric for assessing countries' intrinsic trade capabilities, paving the way for diverse and promising avenues of further research and development. For instance, analyzing the temporal evolution of global mass distribution could shed light on the nuanced dynamics of economic phenomena. Additionally, studying the spatial distribution of mass within mobility networks could unveil the complex landscape of transportation systems.

Beyond these applications, our method has significant potential to extend the scope of gravity models to various other fields, such as supply chain networks, social networks, ecological dynamics, and more. By reducing dependence on external data and allowing a more direct exploration of inherent system properties, our approach significantly broadens investigative horizons in these domains.

In conclusion, our study introduces an innovative self-consistent inference formulation for gravity models, advancing the methodology beyond the limitations of traditional approaches. This advancement refines the gravity modeling paradigm, reducing dependency on external data and enabling a more direct exploration of phenomena within the gravity model framework. The development and application of our approach open new avenues for analyzing complex systems, with the potential to enrich the application of gravity models in diverse contexts—from international trade to social and ecological networks and beyond. Future research can build on this foundation, further enhancing the utility and scope of gravity models in understanding complex human dynamics.

\backmatter

\section{Method}

\subsection{Algorithm Detail and Implementation}
Our numerical algorithm procedure consists of successive stages. In the initial stage of the simulation (at step $n=0$), we establish a baseline where both the outward and inward masses for each node $i$ are set to unity (${\bar m}_{i}^{\rm out}(0) = {\bar m}_{i}^{\rm in}(0) = 1$), and the deterrence function ${\bar Q}(d; 0)$ is uniformly initialized to 1 for all distance values $d$. This initialization serves as a neutral foundation for the iterative dynamics that follow.

During each iteration step, we adjust the mass distributions to explain the flow data based on the spatial dynamics observed up to the previous step $n$, employing two key update formulas:

\begin{equation}
{\bar m}^{\rm out}_{i}(n+1) = \frac{S^{\rm out}_{i}}{\sum_{j } A_{ij}{\bar m}^{\rm in}_{j}(n) {\bar Q}(d_{ij}; n)},
\end{equation}
where $S^{\rm out}_{i}$ represents the total outward flow from node $i$. The inward mass, ${\bar m}^{\rm in}_{j}(n+1)$, for node $j$ follows a similar updating scheme:

\begin{equation}
{\bar m}^{\rm in}_{j}(n+1) = \frac{S^{\rm in}_{j}}{\sum_{i} A_{ij}{\bar m}^{\rm out}_i(n) {\bar Q}(d_{ij}; n)},
\end{equation}
with $S^{\rm in}_{j}$ encapsulating the total inward flow to node $j$. Note that those update processes match the strength of each node with the respective generated strength.

Each iteration includes a critical update to the deterrence function, ${\bar Q}(d_k; n+1)$, reflecting changes in node interactions and spatial relationships at step $n+1$. R. Cazabet also mentioned a similar concept in his work\cite{Cazabet} and suggested an iterative update of the deterrence function, without detailed analysis. This update uses a piece-wise linear approximation based on equal-frequency binning, which divides the range of distances into segments containing equal numbers of node pairs to ensure uniform sampling across varied distances. This binning technique sorts the distance data, $\{d_{ij}\}$, and groups them into predefined bins, each containing an equal number of data points. If the total number of data points is not divisible by the bin size, the last bin is adjusted to include the remainder, ensuring that it contains slightly more data points as necessary. Within each bin, the deterrence function's value, ${\bar Q}(d_k; n+1)$, is updated based on the average of node pairs that fall within the bin's distance range. Specifically, the update formula is given by:

\begin{equation}
{\bar Q}(d_k; n+1) = \frac{1}{N_k} \sum_{(i,j) \in \text{bin } k} \frac{f_{ij}}{{\bar m}^{\rm out}_{i}(n+1){\bar m}^{\rm in}_{j}(n+1)},
\end{equation}
where $N_k$ denotes the number of node pairs within each bin $k$. This methodical approach to updating the deterrence function ensures that the model accurately reflects the underlying spatial and interaction dynamics of the network, facilitating a robust simulation of the system's evolution over time.

Since these updates do not inherently constrain the scales of the variables, we establish their scales through specific criteria. In particular, we scale all $f_{ij}$ by a constant factor to ensure that the total inferred flow matches the total observed flow. The deterrence function ${\bar Q}(d)$ is normalized such that its maximum value is 1. We further adjust the remaining scales by ensuring that the average of all ${\bar m}^{\rm out}_{i}$ equals the average of all ${\bar m}^{\rm in}_{j}$.

\section{Data availability}
The authors declare that all data supporting the findings of this study are included within this published article and its supplementary information files. The raw data used in the analysis are publicly available and can be accessed through the sources cited in the manuscript. The filtered dataset generated during this study is available from the corresponding authors upon reasonable request.

\section{Code availability}

The implementation of the self-consistent gravity model used in this study is publicly available as a Python package, `scgravity`, at the Python Package Index (https://pypi.org/project/scgravity/). The source code is also accessible via GitHub at https://github.com/42megaparsec/scgravity.

\bmhead{Supplementary information}

This file contains supplementary text, supplementary figures 1 – 3, supplementary equations 1 – 3, and supplementary references.

\bmhead{Acknowledgments}
B. J. Kim acknowledges support from the National Research Foundation of Korea (NRF) grant funded by the Korean government (MSIT) (Grant No. 2019R1A2C2089463). D. Lee and H. Kim acknowledge support from the NRF grant funded by the Korean government (MSIT) (Grant No. NRF-2022R1C1C1005856).
H. C. Jeong acknowledge support from the NRF grant funded by the Korean government (MSIT) (Grant No. NRF-2021R1F1A1063238).

\section*{Declarations}

\bmhead{Competing Interests}
The authors declare no competing interests.

\bmhead{Contributions}
D.L., G.K., H.-C.J., and B.J.K. conceived the research and developed the methodology. W.C. and D.L. extracted the data. D.L. performed the simulations, and all authors contributed to analyzing the results. D.L., B.J.K., and W.C. drafted the manuscript under the supervision of G.K., H.-C.J., and H.K.


\end{document}